\documentclass[12pt,epsf,amssymb]{article}
\usepackage{tabularx}
\usepackage{array}
\usepackage{graphics}
\usepackage{graphicx}
\usepackage{epsfig}
\usepackage{amsmath}
\usepackage{amssymb}
\usepackage{citesort}
\makeatletter

\usepackage{verbatim}

\newcommand{\bea}{\begin{eqnarray}}
\newcommand{\eea}{\end{eqnarray}}
\newcommand{\beq}{\begin{equation}}
\newcommand{\eeq}{\end{equation}}

\newcommand{\pdir}{p\kern -5.2pt\raise 0.2ex\hbox {/}}
\newcommand{\vdir}{v\kern -5.75pt\raise 0.15ex\hbox {/}}
\newcommand{\kdir}{k\kern -5.75pt\raise 0.15ex\hbox {/}}
\newcommand{\epsdir}{\epsilon\kern -5.0pt\raise 0.15ex\hbox {/}}
\newcommand{\bvdir}{\bar{v}\kern -5.75pt\raise 0.15ex\hbox {/}}
\newcommand{\Ddir}{D\kern -7.75pt\raise 0.20ex\hbox {/}}
\newcommand{\Adir}{A\kern -7.75pt\raise 0.20ex\hbox {/}}
\newcommand{\ldir}{l\kern -5.0pt\raise 0.2ex\hbox{/}}
\newcommand{\varepsdir}{\varepsilon\kern -5.5pt\raise 0.15ex\hbox{/}}

\newcommand{\taud}{\tau_{\frac{1}{2}}}
\newcommand{\taut}{\tau_{\frac{3}{2}}}

\newcommand{\dlz}{\stackrel{\leftarrow}{D^0}}
\newcommand{\drz}{\stackrel{\rightarrow}{D^0}}

\newcommand{\ra}{\rightarrow}

\makeatother

\begin{document}
\thispagestyle{empty}

\begin{flushright}
\begin{tabular}{l}
{\tt LPT Orsay, 04-43}\\
{\tt SWAT/401}\\
{\tt PCCF RI 0410}\\
\end{tabular}
\end{flushright}
\vskip 2.6cm\par
\begin{center}
{\par\centering \textbf{\LARGE Lattice measurement } }\\
\vskip .45cm\par
{\par\centering \textbf{\LARGE of the Isgur-Wise functions
 $\tau_{1/2}$ and $\tau_{3/2}$ } }\\
\vskip 0.9cm\par
{\par\centering 
\sc  D.~Be\'cirevi\'c$^a$, B.~Blossier$^a$, Ph.~Boucaud$^a$, G.~Herdoiza$^b$,
J.P.~Leroy$^a$, A.~Le~Yaouanc$^a$, V. Mor\'enas$^c$, O.~P\`ene$^a$}
{\par\centering \vskip 0.5 cm\par}
{\par\centering \textsl{$^a$ 
Laboratoire de Physique Th\'eorique (B\^at.210), Universit\'e
Paris XI-Sud,\\
Centre d'Orsay, 91405 Orsay-Cedex, France.} \\
\vskip 0.3cm\par}
{\par\centering \textsl{$^b$
Department of Physics, University of Wales Swansea,\\
Singleton Park, Swansea, SA2 8PP, United Kingdom.} \\
\vskip 0.3cm\par}
{\par\centering \textsl{$^c$
Laboratoire de Physique Corpusculaire \\
Universit\'e Blaise Pascal - CNRS/IN2P3 F-63000 Aubi\`ere Cedex, France. } \\
\vskip 0.6cm\par}
June 18, 2004
\end{center}

\vskip 0.45cm \begin{abstract}
We propose a method to compute 
the Isgur-Wise form factors $\tau_{1/2}(1)$ and  $\tau_{3/2}(1)$
for the decay of $B$ mesons into orbitally excited (P wave) 
$D^{\ast\ast}$ charmed mesons on the lattice in the 
static limit. We also present the result of an exploratory 
numerical simulation
which shows that the signal/noise ratio allows 
for a more dedicated computation. We find
$\tau_{1/2}(1)=0.38(5)$ and  $\tau_{3/2}(1)= 0.53(8)$,
with yet unknown systematic errors. These preliminary 
numbers agree fairly well with theoretical expectation. 
\end{abstract}
\vskip 2.2 cm

\setcounter{page}{1}
\setcounter{equation}{0}

\renewcommand{\thefootnote}{\arabic{footnote}}
\vspace*{-1.5cm}
\newpage
\setcounter{footnote}{0}
\section*{Introduction}

The scalar heavy-light mesons and more generally the 
first orbital excitations $D^{\ast\ast}$
have attracted attention since years and they
still remain somehow mysterious. The recent discovery 
of a $c \bar s$-scalar meson significantly lighter 
than expected has renewed the interest
in these states~\cite{Aubert:2003fg,Abe:2003zm}. There have been several 
lattice studies of this spectrum~\cite{M+P,lewis}
and a recent rather complete one compares quenched
and unquenched~\cite{Green:2003zz} computations. 
Recently the $H^*_0 \to H \pi$ transition (scalar-pseudoscalar-pion)
have also been considered~\cite{McNeile:2004rf,nous}. 

The transitions of the type $B\to D^{\ast\ast} l\nu$ raise
a serious problem. In the infinite mass limit these decays
are described by the Isgur-Wise form factors 
$\tau_{1/2}$ and  $\tau_{3/2}$~\cite{Isgur:jf}. To make a 
long story short, a series of sum 
rules~\cite{LeYaouanc:2000cj}-\cite{LeYaouanc:2002vr}
have been derived from QCD, all indicating that $\tau_{3/2}$
should be significantly larger than $\tau_{1/2}$.
These sum rules relate the  $\tau_j$ form factors, as well
as form factors related to excitations, to derivatives of
 the ground state Isgur-Wise function $\xi$ and allow to bound
  the latter derivatives in an efficient and useful 
  way~\cite{LeYaouanc:2003rn}-\cite{Jugeau:2004ui}.
 Not only does the slope of $\xi$ verify $\rho^2 >3/4$ but also the curvature 
 and even higher derivatives are bound.
 The limit in which $\tau_{1/2}=0$ has been baptised ``BPS" by
 Uraltsev~\cite{Uraltsev:2003ye}-\cite{Uraltsev:2003uq}
  and was proven to provide intersting hints.
  
  However  the theoretical prediction that $\tau_{3/2}^{(0)}> 
  \tau_{1/2}^{(0)}$
  and hence that  the decay $B\to D^\ast_2$ should be significantly
  larger than the $B\to D^\ast_0$ is not verified by 
  experiment~\cite{Abe:2003zm,Anastassov:1997im}. This is
  the `1/2 $>$ 3/2' paradox~\cite{Uraltsev:2004ta}. One might incriminate 
  the corrections to the infinite mass limit. Another possibility 
  could be that the sum rules are fulfilled by higher excitations
  and that the ground state obeys an opposite hierarchy i.e. 
  $\tau_{3/2}^{(0)}< \tau_{1/2}^{(0)}$\footnote{There is no mathematical
  impossibility for the sum rules to be fulfilled with an reversed 
  hierarchy for the ground state, but it does not seem very likely 
  and is not seen in models.}.
  
  To answer to this question one needs to compute directly 
  $\tau_{3/2}^{(0)}$ and $\tau_{1/2}^{(0)}$. 
  Here  we propose a lattice method to do that. We will work in
  the static quark limit, $m_{b,c} \to \infty$,
  with the four vectors $v' = v = (1,0,0,0)$,
   and we will exhibit we will exhibit operators whose matrix elements
   allow to measure these form factors.
  
 This letter is meant to propose this new method  and to make 
 a feasibility study. We do not intend at this stage
 to provide accurate results for these form factors but
 merely to describe the principle of the method 
 and to show with preliminary simulations that there is 
 good hope to make the precision calculation. 
 
\section{Principle of the calculation}

We are concerned with the matrix element 
of an electroweak current between a pseudoscalar 
or vector heavy-light meson $H^{(\ast)}$  and an orbitally 
excited one $H^{\ast\ast}$. However, in the conditions
of the infinite mass limit on the lattice with the heavy quarks
at rest, both in the initial and final state
($v_\mu = v_\mu'$), this matrix element vanishes. 

The way out is to use a series of relations derived in 
ref.~\cite{Leibovich:1997em}. In that paper it has been shown that 
in the case of a matrix element which vanishes 
linearly in the difference $v'-v$, when $v' \to v$, there are
 non-vanishing  forward matrix elements (for $v'=v$) 
 involving the covariant derivative operator $D_\mu$.
  These matrix elements are proportional to the 
 infinite mass limit 
form factors $\taud(1)$ or $\taut(1)$.
 
 Let us summarise their proof using different notations:
 For simplicity we take $v' = (1,0,0,0)$,
 and $v = v' + v_{\perp}$, where 
 $v_\perp$ is  spatial up
 to higher orders in the difference $v'-v$. 
 We assume that for some Dirac matrix $\Gamma_l$
 
 \beq\label{tau1}
\langle H^{\ast\ast}(v') | \bar h(v') \Gamma_l h(v) |  H^{(\ast)}(v)\rangle  =
t_l^{m}v_{\perp\,m} \tau_j(w) + \cdots,
 \eeq 
 where $w\equiv v\cdot v'$,  $j=1/2$, $3/2$, and $l,m=1,3$ are  spatial indices, $t^{l m}$ is a tensor
 which depends on the final state ($H^{\ast\ast}$) and the initial
 state ($H^{\ast}$ or $H$). The dots represent higher orders 
 in  $v'-v$. From translational invariance in the time direction,
\bea \label{D0}
-i \partial_0 \langle H^{\ast\ast}(v') | \bar h(v') \Gamma_l h(v) 
|  H^{(\ast)}(v)\rangle & =& \\
\nonumber
-i \langle  H^{\ast\ast}(v') | \bar h(v') \left[ \Gamma_l \drz +
\dlz \Gamma_l \right] h(v) |  H^{(\ast)}(v)\rangle &= &
t_l^{m}v_{\perp\,m} \tau_j (w)\left (M_{H^{\ast\ast}} - M_{H}\right) + \cdots.
 \eea
The authors of ref.~\cite{Leibovich:1997em} use the 
  field equation: $(v\cdot D) h(v) = 0$, 
  which implies that 
 \beq
 D^0  h(v') = 0,\qquad D^0  h(v) = - (D \cdot v_\perp)  h(v),
 \eeq 
 whence from  (\ref{D0}) 
 \beq\label{voila}
 i \langle  H^{\ast\ast}(v') | \bar h(v') \Gamma_l  (D \cdot v_\perp)  h(v) | 
 H^{(\ast)}(v)\rangle
 = t_l^{m} v_{\perp\,m} \tau_j(w) \left (M_{H^{\ast\ast}} - M_{H}\right) + \cdots,
 \eeq
which has a finite limit when $v_\perp \to 0$ namely
\beq
i \langle  H^{\ast\ast}(v) | \bar h(v) \Gamma_l D^m h(v) | H^{(\ast)}(v)\rangle
= t_l^{m}  \tau_j(1) \left (M_{H^{\ast\ast}} - M_{H}\right). 
\eeq 
Applying eq.~(\ref{tau1}) to the $J=0\; H^\ast_0$ state we get
from ref.~\cite{Isgur:jf}:
\beq\label{tau0}
\langle  H^\ast_0 (v')| A_i | H(v)\rangle \equiv - \tau_{\frac 1 2}(w)
 v_{\perp\, i},
 \eeq
where $A_i$ is the axial current in the spatial direction $i$
and where our states normalisation is $1/\sqrt{2 M}$ times the one
used in ref.~\cite{Isgur:jf}. 
From eq.~(\ref{tau0}) it results that
\beq\label{scalaire}
\langle H^\ast_0(v) | A_i D_j  | H(v)\rangle  =  i \, g_{ij}\left (M_{H^{\ast}_0} -
 M_{H}\right)  \tau_{\frac 1 2}(1). 
\eeq  
Analogously for the $J=2\; H^\ast_2$ state we have
\beq
\langle H^\ast_2 (v')| A_i  | H(v)\rangle  \equiv \sqrt{3} \; 
\tau_{\frac 3 2}(w) \; 
\epsilon^{\ast \, j}_{i} v_{\perp\, j} + \cdots\;,
\eeq
where $\epsilon^\ast_{ij}$ is the polarisation tensor, whence
\beq\label{tenseur}
\langle H^\ast_2(v) | A_i D_j  | H(v)\rangle  = - i \sqrt{3}\left (M_{H^{\ast}_2} - 
M_{H}\right)
\tau_{\frac 3 2}(1) \epsilon^\ast_{ij}\;.
\eeq

\vspace*{3cm}

\section{Lattice calculations}

To compute the matrix elements in eqs.(\ref{scalaire})
and (\ref{tenseur})  on the lattice we first need a discretized 
expression for the covariant derivative. We choose the symmetrised
form
\beq\label{covariant}
D_i(\vec x,t) \to \frac 1{2a}\left (U_i(\vec x,t) - U_i^\dagger(x-
\hat i,t) \right),
\eeq
where $U_i(\vec x,t)$ is the link variable of the lattice.

\subsection{Interpolating fields}
\label{IF}
The interpolating fields for orbitally excited states have been studied 
in ref.~\cite{Lacock:1996vy}. Smearing is used not only to 
improve the signal/noise ratio by 
 better isolating the ground state, but also 
to produce convenient interpolating fields for the $0^-$, $0^+$ and $2^+$
states. Inspired by
 ref.~\cite{Boyle:1999gx} we replace the quark fields $q(x)$ by  
\bea \label{Smearing}
q(x)& \rightarrow &\sum\limits_{r=0}^{R_{max}} {\scriptstyle
(r+\frac{1}{2})}^2 \phi({\scriptstyle r})
 \sum\limits_{i=x,y,z} \Bigg\{
\left[ \prod\limits_{k=1}^r U^F_i({\scriptstyle
x+(k-1)\hat{i}})\right] q({\scriptstyle x+r\hat{i}})
 \;\begin{matrix}{( r \hat i)\; \delta_{il}}\\{1}\end{matrix}\; \nonumber\\ 
 &+& \left[
\prod\limits_{k=1}^r U_i^{F^\dagger}({\scriptstyle
x-k\hat{i}})\right] q({\scriptstyle x-r\hat{i}})
\;\begin{matrix}{(-r \hat i)\; \delta_{il}}\\{1}\end{matrix}\; \Bigg\} \;
, \eea 
where the upper (lower) expressions generate negative (positive) parity
smearing functions.
 The vector $( \pm r \hat i)\; \delta_{il}$
 is  introduced to generate an orbital excitation in the direction $l$. 
The wave function $\phi({r})$ is a radial function chosen to optimise  
the overlap with the ground state.
We take $\phi({ r}) = e^{-r/R_b}$, where $R_b$ is a parameter which is 
fixed by requiring the smearing to be optimal. 
Note that it is not necessary to normalize the wave function since 
the normalisation factors cancel in the computation of matrix elements. 
The smearing also includes the so-called fuzzing, see
  ref.~\cite{Abada:2003un}.
For convenience, we
will use the following notation for the interpolating fields:
\beq
 \{ \bar{h}(x)   \; \begin{matrix}\gamma_k \,{ r}^{l}(x) \\
{\Gamma}\end{matrix} \;q(x+ \vec r)\}\;,
\eeq
where ${ r}^{l}$ indicates the presence of 
$( \pm r \hat i)\; \delta_{il}$ in eq.~(\ref{Smearing}),
 $\vec r$ is a generic vector for the distance between the light 
and heavy quark field and $\Gamma=1$ ($\Gamma=\gamma_5$) for the $0^+$ ($0^-$)
meson.

Using the smeared  quark fields from eq.~(\ref{Smearing})
we now define the interpolating fields. We concentrate 
on the $0^+$ ($2^+$) states which correspond to $j=1/2$ ($j=3/2$).
The $0^+$-state can be described according to two distinct
interpolating fields: 
\beq\label{0p}a)\quad
\bar  h(x) \, q(x+ \vec r) \quad {\rm and}\quad b)\quad 
 \frac 1 {\sqrt 3} \,\bar h(x)\,\left(   \vec \gamma \cdot {\vec r} \right) 
 \, q(x+ \vec r).\quad
\eeq
These two interpolating fields differ in that the Dirac matrix
is diagonal (antidiagonal) for $1$ ($\vec \gamma \cdot {\vec r} $)
inducing the coupling of the heavy quark to the ``small'' (``large'')
 component of the light quark field. The latter is just the quark model
 combination of quark-spin 1 with orbital momentum 1 to generate $J=0^+$. 

Concerning the $2^+$ states, the same duality of interpolating fields exists.
In this letter we only consider the quark-model type interpolating fields. This 
gives the five $J=2$ states~\cite{Lacock:1996vy} which we may write as follows:
\bea\label{2p}
 &a)&\quad - \frac 1 {\sqrt{2}} \bar{h}(x)\,\left[   \gamma_i \cdot { r_j}(x) + \gamma_j
 \cdot { r_i}(x)\right] \, q(x+ \vec r)
,\quad i\ne j, \\ \nonumber
&b)&\quad - \frac 1 {\sqrt{2}}
\bar{h}(x)\,\left[   \gamma_1 \cdot { r_1}(x) - \gamma_2 \cdot { r_2}(x)\right]
\, q(x+ \vec r),
\\ \nonumber
&c)&\quad \frac 1 {\sqrt{6}}
\bar{h}(x)\,\left[   \gamma_1 \cdot { r_1}(x) + \gamma_2 \cdot { r_2}(x)
-2  \gamma_3 \cdot { r_3}(x)\,\right] q(x+ \vec r).
\qquad
\eea
These are, as expected, symmetric traceless tensors.


\subsection{Two-point Green functions and $1/2-3/2$ mass splitting}

The $0^+$ two-point Green function is written as 
\beq
C_{2;0}^{1} =  \left \langle
\sum_{\vec x} {\rm Tr} \left[  P^{\vec 0}_{0,t_x}\,\frac {1+\gamma_0}2
 S(\vec r(0),0;x+\vec r(x);U)
\right] \right \rangle_U, 
\eeq
when we use the interpolating field in eq.~(\ref{0p},a)~\footnote{
The superindex $1$ ($\vec\gamma\cdot \vec r$)
refers to the use of the $a$ ($b$) interpolating field in 
eq.~(\ref{0p}).}. $P^0_{0,t_x}$ 
is a temporal Wilson line~\footnote{Indeed we compute the
 two point function using
 the interpolating fields in eqs.~(\ref{0p}) and~(\ref{2p})
  properly shifted in space so as to have 
 the light propagator ending at the origin.}
 corresponding to the Eichten-Hill action for 
 the static quark~\cite{eichten-hill}:   
\beq
P_{t_x,t_y}^{\vec x}  = \delta({\vec x} -
{\vec y}) \prod\limits_{t_z=t_x}^{t_y-1} U^{\rm hyp}_t(x+t_z\,\hat{t}), \;
\eeq
using hypercubic blocking~\cite{Hasenfratz:2001hp}~-~\cite{DellaMorte:2003mn}.

The two-point Green functions with $\gamma_i r_j$ 
interpolating fields allow an interesting 
comparison between the $j=1/2$ and the $j=3/2$ cases.
They  will contain terms of the general form
\beq\label{2pbis}
C_{2;J}^{ijkl}(0,t_x) =
\left \langle
\sum_{\vec x} {\rm Tr} \left[\gamma_i r_j(0)\, P^{\vec 0}_{0,t_x}\,\frac 
{1+\gamma_0}2
\gamma_k r_l(x) S(\vec r(0),0;x+\vec r(x);U)
\right] \right \rangle_U,  \eeq
where $(i,j)=(k,l)$ for the case of interpolating field~(\ref{2p},a)
 or $i=j, k=l$ for the case~(\ref{2p},b), (\ref{2p},c) and for the $0^+$ case.  
$J$ stands for the total angular  momentum
($J=0,2$). 


 After some simple algebra we can write 
\bea\label{2termes}
-C_{2;J}^{ijkl}(0,t_x) = \left \langle\sum_{\vec x} {\rm Tr}\, P^{\vec 0}_{0,t_x}\, 
\left[\left(\delta_{jl} 
\pm i\epsilon^{jlm}\sigma_m\right) r_j(0) \frac {1-\gamma_0}2
 r_l(x) S(\vec r(0),0;x+\vec r(x);U) 
\right] \right \rangle_U. 
\eea
Let us define $C_{2,\delta(r_j(0),r_j(x))}$ and 
$C_{2,\epsilon^m(r_j(0),r_l(x))}$ respectively
the two terms in eq.~(\ref{2termes}). It is easy to see from the 
second interpolating field in eq.~(\ref{0p}) that
the $0^+$ two point Green function writes as
\bea\label{ls12}
-C_{2;0}^{\vec\gamma\cdot \vec r}
& = &\frac 1 3 \sum_{j=1,3}C_{2,\delta(r_j(0),r_j(x))} + \frac i 3 
\sum_{i,j,k}\left[C_{2,\epsilon^k(r_i(0),r_{j}(x))} +
C_{2,\epsilon^k(r_{j}(0),r_i(x))} \right]  \nonumber \\
&= &C_{2,\delta(r_1(0),r_1(x))} + i
\left[C_{2,\epsilon^3(r_1(0),r_2(x))} 
-C_{2,\epsilon^3(r_2(0),r_1(x))} \right],
\eea
where $i,j,k$ are in cyclic order and where we have taken advantage of 
the hypercubic symmetry in the r.h.s.

Taking now any of the $2^+$ meson interpolating fields
and using the again the cubic symmetry we get 
\beq\label{ls32}
-C_{2;2}=C_{2,\delta(r_1(0),r_1(x))} - \frac i 2
\left[C_{2,\epsilon^3(r_1(0),r_2(x))} 
-C_{2,\epsilon^3(r_2(0),r_1(x))} \right].
\eeq
The difference between the $j$=$1/2$ and the $j$=$3/2$ state
is thus related
to the relative sign and coefficient of the $\epsilon$-term compared
to the direct one. The effective energy  is obtained by 
taking minus the time derivative of the logarithm of the 
two-point function. The energy difference between $j=1/2$ and $j=3/2$
 is thus proportional to 
\beq\label{ls1}
-i \left \langle [r_1(0) \dot r_2(x) -r_2(0) \dot r_1(x)] \;{\rm Tr}
\, P^{\vec 0}_{0,t_x}\, \left[ 
\sigma_3 \frac {1-\gamma_0}2
  S(\vec r(0),0;x+\vec r(x);U) 
\right] \right \rangle_U.
\eeq
In a non-relativistic limit $ \dot {\vec r} = i \vec p/m$. This 
imaginary velocity comes from
the derivation versus the imaginary time. Then the  expression 
in eq.~(\ref{ls1}) is reminiscent of a $LS$-term: $(\vec r \times \vec p )\cdot
\vec  \sigma$ 
except that the operators $\vec r$  and $\vec p$ are not taken at the same time.
$\sigma$ in eq.~(\ref{ls1}) acts on the heavy quark but the trace will make it 
also act on the light quark. It is interesting 
that the coefficients of the last terms in eqs.~(\ref{ls12}) and~(\ref{ls32})
are in the ratio $(1),(-1/2)$ which is exactly the ratio of  the 
$LS$-eigenvalues
for $j=1/2,3/2$, built up from the combination of $L=1$ and $s=1/2$:
\beq\label{alaLS}
2\,<\vec L\cdot \vec S> = j(j+1) - \frac 3 4 -2 =\begin{matrix} -2\\ 1 
\end{matrix}\quad {\rm for}\quad j = \begin{matrix}  1/2\\  3/2 
\end{matrix} \qquad . 
\eeq
From eqs.~(\ref{ls12}) and (\ref{ls32}) it is obvious that  
if these $LS$-type terms did vanish the two-point correlators
$C_{2;0}$ and $C_{2;2}$ would be equal, which 
would then imply $M_{H^\ast_2}= M_{H^\ast_0}$.  In this limit
 the normalisation of 
the interpolating fields in eqs.~(\ref{0p}) and (\ref{2p}) has 
further ensured the equality of the multiplicative constants
${\cal Z}_{2;0}$ and ${\cal Z}_{2;2}$, where the ${\cal Z}_{2;J}$
 are defined from
\bea
 C_{2;J}(t_x) =  ({\cal Z}_{2;J})^2 \, e^{-M_{H^{\ast}_J}\; t_x} \, ,
\eea 
at large time $t_x$.
\subsection{Three-point Green functions and $\taud$-$\taut$ splitting}
The three point Green functions of the axial current using 
interpolating fields with $ \gamma_i r_j$  will 
contain terms  of the general form
\bea\label{3points}
&&C_{3,JA_k5}^{ijkl}(0,t_y,t_x)=\frac 1 2
\Bigg \langle
\sum_{\vec x,\vec y } {\rm Tr} \Big[\gamma_i r_j(0) \, P^{\vec 0}_{0,t_y}\,
\frac {1+\gamma_0}2
\gamma_k \gamma_5 \\ \nonumber
& &\left \{U_l(0,t_y) P_{t_y,t_x}^{\hat l} 
S(\vec r(0),0;x+\vec r(x)+\hat l;U) - U_l^\dagger(-\hat l,t_y) P_{t_y,t_x}^{-\hat l}
S(\vec r(0),0;x+\vec r(x)-\hat l;U)\right \}\gamma_5\Big] \Bigg \rangle_U \;,
\eea
where we have used eq.~(\ref{covariant}) in units of $a$.
  Writing for short the term in the curly bracket as $D_l(y)\cdots\gamma_5$,
and we see can write 
\beq\label{3points2}
- C_{3,JA_k5}^{ijkl}(0,t_y,t_x)=\frac 1 2\left \langle 
\sum_{\vec x,\vec y }{\rm Tr} 
\left [\left(\delta_{jl} \pm i\epsilon^{jlm}\sigma_m\right)
\gamma_5 r_j(0) \, P^{\vec 0}_{0,t_y}\,\frac {1+\gamma_0}2 D_l(y)\cdots \gamma_5
\right ]\right\rangle_U \;,
\eeq
where either $(i,j)=(k,l)$ for the $0^- \ra 2^+$ 
transition,  eq.~(\ref{2p},a), or $i=j, k=l$ for the
$0^-\ra 0^+$ one and $0^- \ra 2^+$ with eqs.~(\ref{2p},b) and (\ref{2p},c).
Let us define $C_{3,\delta(r_j(0),D_j(x))}$ and 
$C_{3,\epsilon^m(r_j(0),D_l(x))}$, respectively
the two terms in eq.~(\ref{3points2}). 
From the  interpolating field in eq.~(\ref{0p},b), using
the fact that $i=j$ and $k=l$, and choosing for simplicity
$k=3$, one can derive, 
\bea\label{3ls12}
-C_{3;05}^{\vec\gamma\cdot \vec r}
& = &\frac 1 {\sqrt{3}} C_{3,\delta(r_3(0),D_3(x))} + \frac i {\sqrt{3}} 
\left[C_{3,\epsilon^1(r_2(0),D_{3}(x))} +C_{3,\epsilon^2(r_1(0),D_{3}(x))}
 \right] \;.
\eea
The axial matrix element is then given in Euclidean metric by
\bea
<H^\ast_0|A_3 D_3|H> = \frac{{\cal Z}_{2;0}\;{\cal Z}_{2;5}\;C_{3;05}(0,t_y,t_x)}
{C_{2;0}(0,t_y)\;C_{2;5}(0,t_x-t_y)} = \left (M_{H^{\ast}_0} -
 M_{H}\right)  \tau_{\frac 1 2}(1) \; ,
\eea
where we have used eq.~(\ref{scalaire}).
This leads, using cubic symmetry, to
\bea\label{tau12}
\left (M_{H^{\ast}_0} -
 M_{H}\right) \;\sqrt 3 \;\tau_{\frac 1 2}(1) = {\cal Z}_{2;0}\;{\cal Z}_{2;5}
  \frac{C_{3,\delta(r_3(0),D_3(x))}
 + 2\,i \,C_{3,\epsilon^1(r_2(0),D_{3}(x))}
 }{C_{2;0}(0,t_y)\;C_{2;5}(0,t_x-t_y)}\;,
\eea
plus all terms deduced by cubic symmetry.

From eq.~(\ref{tenseur}), in Euclidean metric,  we get
\bea\label{a1d2}
<H^\ast_2|\frac {A_1 D_2 + A_2 D_1}2|H> = \frac{{\cal Z}_{2;2}\;{\cal
Z}_{2;5}\;C_{3;25}(0,t_y,t_x)}
{C_{2;2}(0,t_y)\;C_{2;5}(0,t_x-t_y)} =\sqrt{\frac 3 2}\; \left (M_{H^{\ast}_2} -
 M_{H}\right)\; \tau_{\frac 3 2}(1)\,, 
\eea
where we have used the polarisation tensor
\beq  \epsilon = 
\begin{matrix}  0 &\frac 1 {\sqrt 2} &0 \\  \frac 1 {\sqrt 2} &0 &0 \\ 0&0 &0\\
 \end{matrix}\; ,
\eeq 
for the $2^+$ state in eq.~(\ref{2p},a) with $(i,j)=(1,2)$.
From the interpolating field in eq.~(\ref{2p},a) using
the fact that $(i,j)=(k,l)=(1,2)$   one can derive, 
\bea\label{3ls32}
-C_{3;25}
& = &-\frac 1 {2\,\sqrt{2}} \sum_{l=1,2}C_{3,\delta(r_l(0),D_l(x))}
 + \frac i {2\,\sqrt{2}} 
\left[C_{3,\epsilon^3(r_1(0),D_{2}(x))} +C_{3,\epsilon^3(r_2(0),D_{1}(x))}
 \right]\;,
\eea 
and using cubic symmetry and eq.~(\ref{a1d2}) 
\bea\label{tau32}
\left (M_{H^{\ast}_2} -
 M_{H}\right)\; \sqrt 3 \;\tau_{\frac 3 2}(1)  = 
 {\cal Z}_{2;2}\;{\cal Z}_{2;5}\;
 \frac{ C_{3,\delta(r_1(0),D_1(x))}
 -i \,C_{3,\epsilon^1(r_2(0),D_{3}(x))} 
}{C_{2;2}(0,t_y)\;C_{2;5}(0,t_x-t_y)}\;.
\eea
It can be checked that all the other states in eq.~(\ref{2pbis}) lead to
the same formula~(\ref{tau32}) up to a cubic rotation.  The numerators in
equations~(\ref{tau12}) and (\ref{tau32}) exhibit identical $C_{3,\delta}$ terms
and $C_{3,\epsilon}$, the latter differing only by  multiplicative
 coefficients which,
 once more, are proportional to the $LS$ eigenvalues given in eq.~(\ref{alaLS}).
Combining the results of eqs.~(\ref{ls12}), (\ref{ls32}), (\ref{tau12}) and
(\ref{tau32}) we may conclude that, if the $C_{2,\epsilon}$ and 
$C_{3,\epsilon}$ terms did vanish, we would
 get $M_{H^\ast_2}=M_{H^\ast_0}$ and $\tau_{3/2} = \tau_{1/2}$.  
\section{Condition and results of the simulation}

\begin{figure}[htbp]
\vspace*{-.1cm}
\begin{center}
\begin{tabular}{@{\hspace{-0.7cm}}c}
\epsfxsize10.0cm\epsffile{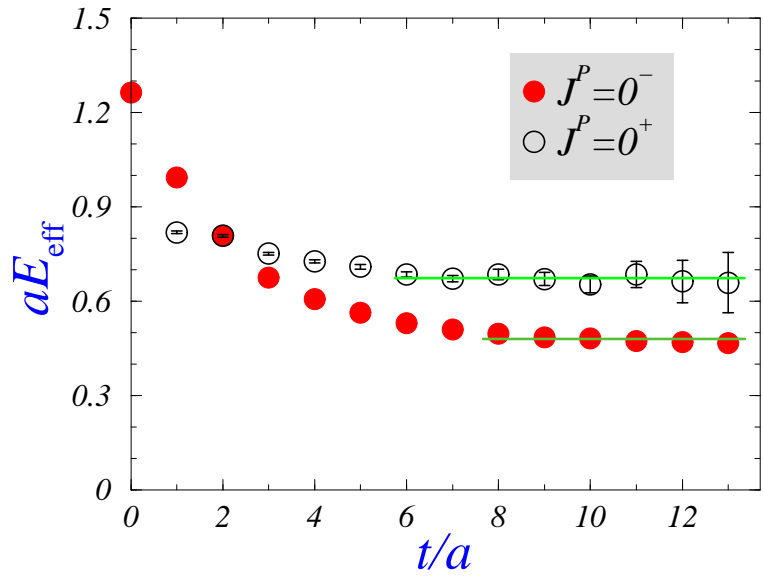}   \\
\end{tabular}
\caption{\label{masseffs0p5}{\small\sl Signals for the effective 
binding energies for 
the pseudoscalar and the scalar heavy-light mesons.}}
\end{center}
\end{figure}

\begin{figure}[htbp]
\vspace*{-.1cm}
\begin{center}
\begin{tabular}{@{\hspace{-0.7cm}}c}
\epsfxsize10.0cm\epsffile{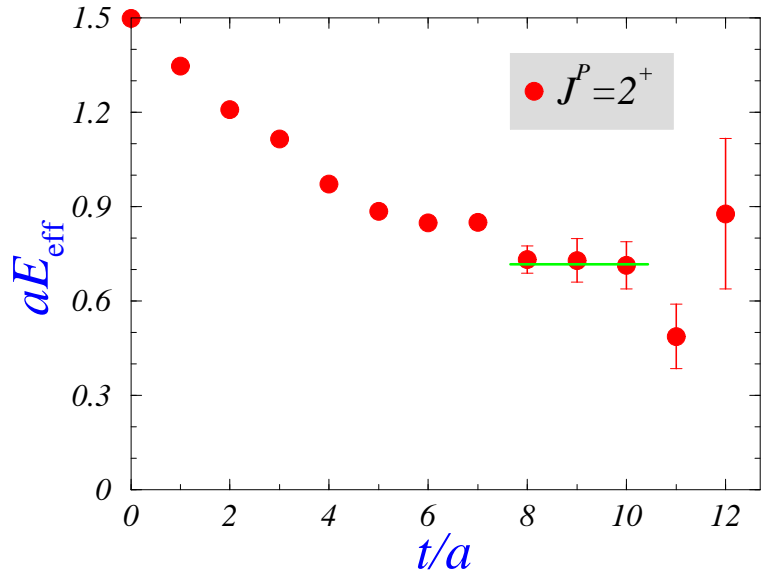}   \\
\end{tabular}
\caption{\label{masseff2plus}{\small\sl Signal for the effective 
binding energies $E_{\rm eff}$ for 
the $2^+$ heavy-light mesons.}}
\end{center}
\end{figure}

Results presented for Isgur-Wise functions $\taud(1)$ and $\taut(1)$ 
are obtained from the quenched simulation on a $16^3 \times 40$ 
lattice at $\beta=6.0$. We collected
$580$ independent $SU(3)$ gauge configurations in the quenched approximation 
using the non perturbatively $\mathcal{O}(a)$ improved Wilson fermion action with 
$C_{SW}=1.769$. The light-quark propagator is computed with the hopping parameter 
$\kappa=0.1334$, which corresponds to a pseudoscalar ``light-meson" mass of 800 MeV. 
For the static quark we use the ``hyp" links as written previously.
In fig.~\ref{masseffs0p5} we plot the binding energy for the scalar and
 the pseudoscalar 
meson and in fig. \ref{masseff2plus} we plot the binding energy for the $2^+$ heavy-light
meson. 

The scalar meson has been computed using the interpolating 
field~\footnote{This choice shows
up a better signal than the one using eq.~(\ref{0p},b). A comparison of these
signals has been performed in \cite{nous}.}
in eq.~(\ref{0p},a): $\bar  h(x) \, q(x+ \vec r)$. The 
tensor meson has been
computed using the properly averaged interpolating fields  in eq.~(\ref{2p}).
We get  $\Delta \equiv m_{H^\ast_0}-m_H =400(12) - 411(16)$ MeV at 
$\beta=6.0$. Only statistical errors are considered.
It agrees reasonably with ref.~\cite{Green:2003zz} where 
$\Delta\sim 400 (40)$ MeV.
Our present signal for the tensor-meson effective mass is still
very poor, 
  $m_{H^\ast_2}-m_{H} = 0.50(8)$ GeV, which leads to 
  $m_{H^\ast_2}-m_{H^\ast_0} = 0.10(8)$ GeV. 
  The large relative error reflects the poor quality 
  of the plateau in fig.~\ref{masseff2plus}. 
  Clearly a more refined simulation is needed here. In particular we have
not yet optimised the wave function for the smearing of the tensor meson.
Our result agrees with the result of ~\cite{Green:2003zz} where 
  we read from table 2 ($Q_3$)
  :  $m_{H^\ast_2}-m_{H} = 0.48(2)$ GeV, 
   and   $m_{H^\ast_2}-m_{H^\ast_0} = 0.08(4)$ GeV.
   
  Experimentally the situation is not yet clear:
whereas  Belle~\cite{Abe:2003zm} reported 
  $m_{D^\ast_2} -  m_{D^\ast_0}= 153(36)$ MeV~\footnote{Note however
  than in Belle experiment the narrow and broad $J^P=1^+$ resonances,
  usually interpreted as $j=3/2$, $j=1/2$ respectively, are practically
  degenerate in mass: $m_{D_1^0}= 2421(2)$ MeV and $m_{D_1^{'0}}= 2427(50)$  },  
  FOCUS~\cite{Link:2003bd} finds $m_{D^\ast_2} -  m_{D^\ast_0}= 61(41)$ MeV.
  Anyway large $1/m_c$ corrections are expected. 

\begin{figure}[h]
\vspace*{1.0cm}
\begin{center}
\begin{tabular}{@{\hspace{-0.7cm}}c}
\epsfxsize10.0cm\epsffile{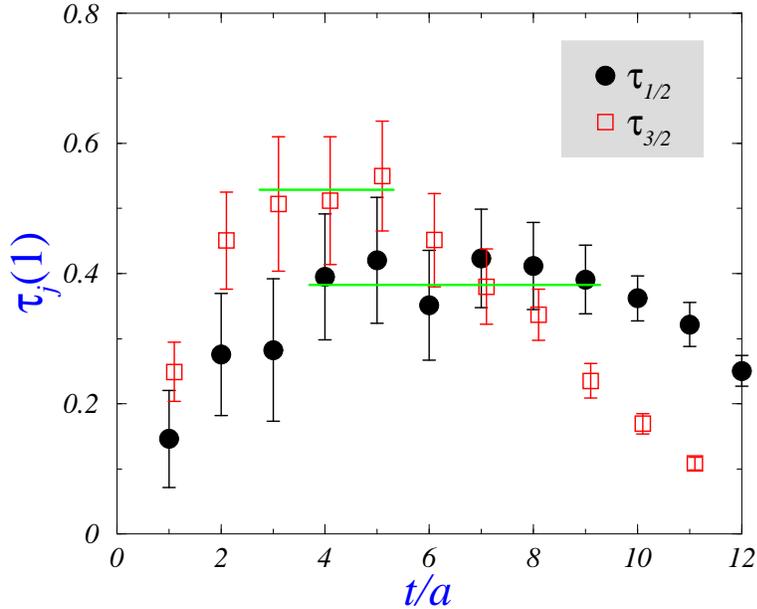}   \\
\end{tabular}
\caption{\label{tauplat}{\small\sl Signals for the ratios defined in
eq.~(\ref{tau}); 
from the fit in $t/a \in [4,9]$ and $t/a \in [3,5]$ 
respectively, we obtain the value of $\taud$ and $\taut$}.}
\end{center}
\end{figure}

In fig ~\ref{tauplat}, we  plot the ratios 
\bea\label{tau}
\tau_{\frac 1 2}(1)&=&\frac 1 {(M_{H^{\ast}_0} - M_{H})} 
\frac{{\cal Z}_{2;0}\;{\cal Z}_{2;5}\;C_{3;05}(0,t_y,t_x)}
{C_{2;0}(0,t_y)\;C_{2;5}(0,t_x-t_y)} \;,\\ \nonumber
 \tau_{\frac 3 2}(1)&=&
\sqrt{\frac 2 3} \frac{1}{(M_{H^{\ast}_2} - M_{H})} 
\frac{{\cal Z}_{2;2}\;{\cal Z}_{2;5}\;C_{3;25}(0,t_y,t_x)}
{C_{2;2}(0,t_y)\;C_{2;5}(0,t_x-t_y)}\;,
\eea 
 where the source operator has been fixed at $t_x=13 a$.
 The equality is valid on the plateau.  

\section{Results, Discussion and Conclusions}

In this letter we propose a method to compute on the lattice, 
at the infinite mass limit, the zero recoil Isgur-Wise form factors  
$\taud(1)$ and $\taut(1)$  relevant to the decay of a
heavy pseudoscalar meson into orbitally excited states. 
The main feature of the method is
contained in eqs.~(\ref{scalaire}) and (\ref{tenseur}).
It uses matrix elements of the axial current multiplied 
by covariant derivatives. 

We have also performed an exploratory lattice study in order to estimate 
if this method is practically usable. We find that 
 the signal/noise ratio is encouraging if one considers
that there is still room for improvement.

Our results are
\beq\label{result}
\taud(1) = 0.38(4)(?)\quad{\rm and}\quad \taut(1) = 0.53(8)(?)\;,
\eeq
where the question mark represent yet unknown systematic errors.
 We have also:
\beq
\taut(1)^2 - \taud(1)^2 = 0.13(8)(?)
\eeq 
 Within $1.5 \sigma$ it saturates 
the Uraltsev sum rule~\cite{Uraltsev:2000ce}: 
$\sum_n|\taut^{(n)}(1)|^2 - |\taud^{(n)}(1)|^2 = \frac 1 4$.
Note that an approximate saturation of the sum rule
 by the ground states is seen in several
  models~\cite{Morenas:1997nk,Cheng:2003sm}
 although there  is no strong theoretical reason for that.

The result for $\taud(1) = 0.38(4)(?)$ is
 presumably more reliable than the 
one on $\taut(1)$ since the two-point signal for the $0^+$ meson
is much better  than the one 
for the $2^+$ meson, see figs.~\ref{masseffs0p5} and \ref{masseff2plus}. 
Our result for $\taud(1)$ is somewhat larger than the predictions of 
the  covariant quark models a la Bakamjian-Thomas (BT)~\cite{Morenas:1997nk} 
which predict,
 for the preferred potentials~\footnote{It should be stressed 
 that the BT method provides a framework in which 
 different potentials can be used. The physics prediction 
 depends, of course, on the chosen potential.}, $\taud(1) \in 
 [0.1 , 0.23]$ and $\taut(1)\in  [0.43,0.54]$.
The latter agrees well with eq.~(\ref{result}). Both numbers
of eq.~(\ref{result}) are compatible with a recent calculation
  based on a covariant light-front approach with simple 
 harmonic oscillator wave functions which are not derived
 from a potential~\footnote{ The  covariant light-front
framework is equivalent, in the infinite mass limit, to the BT
one. The practical
predictions  depend, however, on the chosen parameters
and shape of the wave function. We worry if the use of gaussian
wave functions, which is frequent, is a good one as it neglects 
the short distance potential.}: $\taud(1)=0.31$ and 
$\taut(1)=0.61$~\cite{Cheng:2003sm}. It is interesting to not that 
both in ref.~\cite{Morenas:1997nk}
and in~\cite{Cheng:2003sm}, 
$\taut(1)-\taud(1)\simeq 0.3$,  which might be a 
general feature of BT covariant quark models (see eq.~(5.1) in 
ref.~\cite{Morenas:1997nk}). This is somewhat larger than the difference 
between central values of~(\ref{result}). We also agree with  an older QCD 
sum rule estimate~\cite{Colangelo:1998ga}:
 $\taud(1) = 0.35(8)$.

The quantity we compute on the lattice is a physical quantity: the product of
a mass difference times the form factors $\taud(1)$ and $\taut(1)$. We thus 
expect no multiplicative renormalisation to be needed. A closer scrutiny of
this question is underway, in particular to understand if the hypercubic
treatment of the Wilson line can have some effect on the discretized 
covariant derivative we use. Of course a complete control of 
systematic effects is also needed: finite volume, mass of the light quark,
finite lattice spacing. In section \ref{IF} a duality of interpolating
fields has been pointed out. A systematic comparison of the their predictions 
is still missing.

\section*{Acknowledgement}

\hspace*{\parindent} We  thank Nikolai Uraltsev for illuminating discussions.
The simulation was performed with the APE1000 located in the Centre
 de Ressources Informatiques (Paris-Sud, Orsay) and purchased
thanks to a funding from the Minist\`ere de 
l'Education Nationale and the CNRS.


\end{document}